\documentstyle[prl,aps,amsmath,amssymb,epsfig,twocolumn]{revtex}

\begin{document}
\draft
\title{A covering property of Hofstadter's butterfly}
\author{R.~Ketzmerick,$^{1,2}$ K.~Kruse,$^1$ F.~Steinbach,$^1$ and  
T.~Geisel$^1$}

\address{
$^1$Max-Planck-Institut f\"ur Str\"omungsforschung und Institut f\"ur 
Nichtlineare Dynamik der Universit\"at G\"ottingen, Bunsenstr. 10, D-37073
 G\"ottingen, Germany \\
$^2$Physics Department and Institute for Theoretical Physics, UCSB, 
Santa Barbara, California 93106}
\date{\today}
\maketitle
\begin{abstract}
Based on a thorough numerical analysis of
the spectrum of Harper's operator,
which describes, e.g., an electron on a two-dimensional lattice subjected 
to a magnetic field perpendicular to the lattice plane, we make the 
following conjecture: For any value of the incommensurability parameter 
$\sigma$ of the operator its spectrum can be covered by the bands of the 
spectrum for every rational approximant of $\sigma$ after stretching them 
by factors with a common upper bound. We show that this conjecture has the
following important consequences: For {\it all} irrational values of 
$\sigma$ the spectrum is (i) a zero measure Cantor set and has (ii) a 
Hausdorff dimension less or equal to 1/2. We propose that our numerical 
approach may be a guide in finding a rigorous proof of these results.
\end{abstract}
\pacs{PACS numbers: 73.20.Dx, 71.30.+h, 02.30.Tb }
\narrowtext

\section{ Introduction}

The Harper model\cite{harper} may be considered as the most important 
system in the realm of quasi-periodic systems. It is described by the 
one-dimensional tight-binding Hamiltonian
\begin{equation}
H_{\lambda,\sigma,\nu} = \sum_{n}\,a^{\dagger}_{n+1} a^{ }_{n}+
a^{\dagger}_{n-1}a^{ }_{n}+\lambda \cos (2\pi\sigma n+\nu) 
a^{\dagger}_{n}a^{ }_{n},
\label{haop}
\end{equation}
where $\lambda$, $\sigma$, and $\nu$ are real valued parameters, and 
$a^{\dagger}_{n}$ and $a^{ }_n$ are creation and annihilation operators at
site $n$. Depending on whether $\sigma$ is rational or irrational this 
Hamiltonian is periodic or quasi-periodic, resp. The model exhibits rich
spectral features which showed up for the first time in the numerical work
of Hofstadter\cite{hofst} who computed its spectrum for $\lambda=2$ as a
function of $\sigma$. The beautiful resulting graph (Fig.~1) is today 
known as Hofstadter's butterfly and has attracted a lot of attention among
physicists and mathematicians ever since.

From a physical point of view the Harper model was first introduced to 
describe electrons moving in a two-dimensional periodic potential with a 
square unit cell and subjected to a weak magnetic field perpendicular to 
the potential plane\cite{harper,hofst,azbel,langbein}. In this case, 
Harper's model describes the eigenfunctions in one direction of the 
underlying Bravais lattice after separating plane waves in the 
perpendicular direction. Then, the parameter $\lambda$ is twice the ratio 
of the modulation amplitudes in these two directions, $\sigma$ is given by
the number of magnetic flux quanta per unit cell of the potential, and 
$\nu$ is the wave number of the plane wave being separated. Such a system 
can be realized, e.g., by lateral superlattices on semiconductor 
heterostructures in which the first experimental indications of 
Hofstadter's butterfly were found recently\cite{schloess}. Harper's model 
also appeared in the opposite limit, namely for electrons in a 
two-dimensional periodic potential with strong magnetic field\cite{rauh},
as well as in other physical contexts, e.g., phonons in a quasi-periodic 
potential\cite{burk} and superconducting networks\cite{ram}. Apart from 
describing real physical systems it is an important and simple model 
exhibiting a metal-insulator transition\cite{aubry}: if $\sigma$ is a 
typical irrational, then for $\lambda<2$ all eigenfunctions are extended 
while for $\lambda>2$ they are localized. Hence, at $\lambda=2$ a 
metal-insulator transition occurs and unusual properties of the spectrum 
and the eigenfunctions are expected. This has led mathematicians to study 
Harper's model as a simple example of a system with a possibly singular 
continuous spectrum. In particular, at $\lambda=2$ the spectrum has been 
conjectured to be a zero measure Cantor set for irrational values of 
$\sigma$\cite{hofst,bs82}. This has been shown for a large class of 
irrationals by Helffer and Sj\"ostrand\cite{helff} and Last\cite{la}, 
which, however, do not cover the irrationals typically used in
numerical investigations, namely irrationals with bounded continued 
fraction expansions, e.g., the Golden Mean $(\sqrt{5}-1)/2$. So still the 
question remains whether the spectrum is a zero measure Cantor set for 
{\it all} irrational values of $\sigma$~\cite{lajit,jit}.
 
For both, physicists and mathematicians, the Harper model is also an 
important example for the study of the relation between the spectral 
properties and the dynamics of quantum systems
\cite{abe,geis1,guarn,wilkinsonaustin94,la2,piechon96,geis2}. 
In 1988, Hiramoto and Abe numerically discovered anomalous diffusive 
spreading of wave packets in the Harper model for $\lambda =2$\cite{abe}. 
Furthermore, the model has been used for testing relations proposed 
between the fractal properties of the spectral measure and the temporal 
decay of autocorrelations\cite{geis1} as well as for the asymptotic growth
of the moments of a wave packet\cite{wilkinsonaustin94,piechon96,geis2}.
In these studies one is thus interested in fractal properties of the
spectrum, the easiest being the Hausdorff
dimension. First, it was investigated numerically\cite{wilkinsonaustin94,tang,wi1,bell} and recently a perturbative
calculation has been presented for $\sigma$ the Golden and the Silver 
Mean\cite{fred}. For a class of irrationals, which has zero Lebesgue 
measure, Last proved the Hausdorff dimension to be bounded from above by 
1/2\cite{la}. It remains open whether this bound is valid for {\it all} 
irrationals.

In this article we study the operator~(\ref{haop}) at the critical point 
$\lambda =2$. Based on a numerical analysis we present our conjecture that
there is a constant $R$ such that the spectrum for any value of $\sigma$ 
can be covered by the $q$ bands of the spectrum for any rational 
approximant $p/q$ of $\sigma$ after stretching each of these bands by a 
factor smaller than $R$. This conjecture permits us to prove for {\it all}
irrational $\sigma$ that the spectrum is (i) a zero measure Cantor
set and has (ii) a Hausdorff dimension less or equal to 1/2. After
presenting the relevant basic facts on Harper's operator
in Sec.~II, in Sec.~III we precisely state our conjecture and deduce the
above mentioned statements 
on the measure of the spectrum and its Hausdorff dimension.
Section~IV contains a thorough numerical analysis of Hofstadter's butterfly
substantiating the validity of our conjecture.

\section{ Basic facts}

In the following we review the facts about the spectrum of Harper's 
operator [Eq.~(\ref{haop})] that are used in
the proofs presented in Sec.~III. The relevant energy spectrum
$S(\sigma)$ of the Harper model for $\lambda=2$ and a fixed value of 
$\sigma$ is given by the set
\begin{equation}
S(\sigma) \equiv \bigcup_{\nu} Spec\left(H_{2,\sigma,\nu}\right) ,
\end{equation}
where $Spec\left(H_{2,\sigma,\nu}\right)$ denotes the spectrum of 
$H_{2,\sigma,\nu}$. Figure~1 shows $S(\sigma)$ vs. $\sigma$, the so-called
Hofstadter butterfly~\cite{hofst}, revealing its symmetries with respect 
to $E=0$ and $\sigma=1/2$.

For rational $\sigma=p/q$, with $p$ and $q$ relatively prime, the 
operator~(\ref{haop}) is periodic and $S(\sigma)$ consists of $q$ bands. 
These bands are separated by $q-1$ gaps in the case of $q$ odd and $q-2$ 
gaps for $q$ even, for which the middle bands touch each other at
$E=0$\cite{vMouche}. The Lebesgue measure $|S(p/q)|$ of the spectrum 
obeys\cite{la}
\begin{equation}
\label{last}
\frac{2(\sqrt{5}+1)}{q}<\left|S\left(\frac{p}{q} \right)\right| < 
\frac{8e}{q} ,
\end{equation}
where $e=\exp(1)=2,71\ldots$

For irrational values of $\sigma$, the operator~(\ref{haop}) is 
quasi-periodic and its spectrum contains no isolated points as follows 
from general results on ergodic Jacobi matrices~\cite{sfk}. Furthermore, 
it is independent of $\nu$\cite{sfk}, such that $S(\sigma)$ is a closed 
set. Thus, if $S(\sigma)$ has measure zero it will be nowhere dense and 
therefore will be a Cantor set. As the measure $|S(p/q)|$ of rational 
approximants $p/q$ of $\sigma$ decreases to zero with increasing $q$ 
[Eq.~(\ref{last})], one may naively think that indeed $|S(\sigma)|=0$ 
holds. {\it Rigorously}, however, $|S(\sigma)|$ can only be shown to vanish
for two classes of irrationals $\sigma$: One class consists of the 
irrationals for which all coefficients in the continued fraction expansion
are larger than some constant\cite{helff}, while the irrationals in the 
other class have an unbounded continued fraction expansion\cite{la}. 
The latter proof uses a continuity property of the spectrum, namely that
the spectrum $S(\sigma)$ is H{\"o}lder continuous of order 1/2\cite{avms},
i.e., for every eigenvalue $E$ in the spectrum for $\sigma$ there is a 
nearby eigenvalue $E'$ in the spectrum for $\sigma'$ with
\begin{equation}
\big|E-E'\big|\le 6(2|\sigma-\sigma'|)^{1/2} .
\label{holder}
\end{equation}
This allows to construct a sequence of covers of $S(\sigma)$ using the
approximants $\{p_n/q_n\}_{n=0}^{\infty}$ of the
continued fraction expansion of $\sigma$.
For each $n$ one attaches intervals of length 
$6(2|\sigma-p_n/q_n|)^{1/2}$ to each of the $2q_n$ band edges of 
$S(p_n/q_n)$. According to Eq.~(\ref{holder}) this defines a cover 
$S_{q_n}(\sigma)$ of $S(\sigma)$ consisting of $q_n$ intervals. The 
measure of any of these covers is 
\begin{eqnarray}
S_{q_n}(\sigma) & = & \left|S\left(\frac{p_n}{q_n}\right)\right|+
12q_n\left(2\left|\sigma-\frac{p_n}{q_n}\right|\right)^{1/2} \\ 
		& < & \frac{8e}{q_n}+12q_n\left(2\left|\sigma-
\frac{p_n}{q_n}\right|\right)^{1/2}
\end{eqnarray}
by Eq.~(\ref{last}). For $n\to\infty$ the measure vanishes only if 
$\lim_{n\to\infty}q_n^2 |\sigma-p_n/q_n|=0$,
which is fulfilled for irrationals with an unbounded continued fraction
expansion and proves a zero measure Cantor set spectrum for these
irrationals~\cite{la}.
For general $\sigma$, which fulfill
\begin{equation}
\left| \sigma - \frac{p}{q} \right| \le \frac{C}{q^2} ,
\label{absch}
\end{equation}
for some constant $C>0$, covers as described above 
do not permit to show $|S(\sigma)|=0$.
With respect to the Hausdorff dimension these covers have been used by
Last~\cite{la} to show an upper bound of 1/2 for the zero measure set of
irrationals fulfilling $q_n^4 |\sigma-p_n/q_n| < \tilde{C}$ for all $n$ 
and some constant $\tilde{C}>0$. 
In the following, we will present a conjecture, stating that there is a
useful cover obtained by stretching the bands of a rational approximant of
$\sigma$. Assuming this conjecture  we prove that the Lebesgue measure of 
the spectrum 
is zero as well as that its Hausdorff dimension is less or equal to 1/2 
for {\it all} irrationals. 

\section{ The Conjecture and its Implications} 

In this section we will show how the zero measure Cantor structure of 
spectra for irrational values of 
$\sigma$ and an upper bound of their Hausdorff dimension follow from our 
\newline
{\bf Conjecture~1 :} {\it There is a constant $R>0$ such that for any 
irrational $\sigma\in [0,1]$ and any two rational approximants $p/q$
and $p'/q'$ with $q<q'$, which are obtained by truncating the continued 
fraction expansion of $\sigma$, there 
exists a cover $S_q(p'/q')$ of $S(p',q')$ consisting of $q$ intervals with
the property}
\begin{equation}
\left| S_q \left(\frac{p'}{q'} \right) \right| \le R \,\left| S \left(\frac{p}{q} \right) \right| .
\label{conj}
\end{equation}
Before giving the explicit construction of such a cover and numerically 
substantiating this Conjecture in the
next section, we will now discuss its consequences. One can replace 
$p'/q'$ in Eq.~(\ref{conj})
by an irrational $\sigma$ as is stated in \newline
{\bf Corollary~1 :} {\it Assuming Conjecture 1, there is a constant $R>0$ 
such that for any irrational 
$\sigma\in [0,1]$ and any rational approximant $p/q$ obtained by 
truncating the continued fraction expansion 
of $\sigma$ there exists a cover $\tilde{S}_q(\sigma)$ of $S(\sigma)$
consisting of $q$ intervals with the property}
\begin{equation}
\bigg| \tilde{S}_q(\sigma) \bigg| < \frac{8eR+1}{q}.
\label{corr}
\end{equation}
{\it Proof}: Let $\sigma$ be an irrational and let $p/q$ be a rational 
approximant of $\sigma$, which is 
obtained by truncating its continued fraction expansion. Then, one can
always choose
a sufficently high rational approximant $p'/q'$ of $\sigma$, which again 
is obtained by truncating 
its continued fraction expansion, such that
\begin{equation}
q' \ge 12 q^2 \sqrt{2C},
\end{equation}
where $C$ is the constant of Eq.~(\ref{absch}). We now define a $q$-cover 
$\tilde{S}_q(\sigma)$ of the spectrum for $\sigma$ using the continuity 
property Eq.~(\ref{holder}). To this end we take the $q$ intervals
of the cover $S_q(p'/q')$, as given by Conjecture 1, and enlarge them by
$6(2|\sigma-(p'/q')|)^{1/2}$ at the lower and upper ends, respectively. 
For the Lebesgue measure
$| \tilde{S}_q(\sigma)|$ of $\tilde{S}_q(\sigma)$ we thus have
\begin{eqnarray}
\bigg| \tilde{S}_q(\sigma) \bigg| & \le & \left| S_q
\left(\frac{p'}{q'}\right)\right| + 2 q \cdot 6 \left(2 \left| \sigma-
\frac{p'}{q'} \right| \right)^{1/2} \\
 & \le & \left| S_q\left(\frac{p'}{q'}\right)\right| + 12\sqrt{2C}\;
\frac{q}{q'} \\
 & \le & \left| S_q\left(\frac{p'}{q'}\right)\right| + \frac{1}{q} ,
\end{eqnarray}
where we have used Eq.~(\ref{absch}).
Finally, using Conjecture 1 and Eq.~(\ref{last}) we obtain 
Eq.~(\ref{corr}).
$\qquad \Box$ 

From this Corollary follows \newline
{\bf Theorem~1 :} {\it Assuming Conjecture 1, for any irrational 
$\sigma\in [0,1]$ the set $S(\sigma)$ is of zero Lebesgue measure.}
{\it Proof}: By taking the limit $q\to\infty$ in Eq.~(\ref{corr}) this 
follows immediately.
$\qquad \Box$

The Cantor structure of the spectrum for irrational $\sigma$ is a direct 
consequence of Theorem~1
and stated in \newline 
{\bf Corollary~2 :} {\it Assuming Conjecture 1, for any irrational 
$\sigma\in [0,1]$ the set $S(\sigma)$ is a Cantor set.}
{\it Proof}: Since the spectrum $S(\sigma)$ is closed, Theorem 1 shows 
that it is nowhere dense. As it is also
known to have no isolated points\cite{sfk}, it follows that $S(\sigma)$ is
a Cantor set.$\qquad \Box$

As a second application of Conjecture 1, we now give an upper bound on the
Hausdorff dimension $D_H$ of the 
spectrum. Most of the proof is contained in \newline
{\bf Lemma~2}{ (Ref.~\cite{la})}{\bf :} {\it Let $S\subset{\mathbb R}$, 
and suppose that $S$ has a sequence of covers: 
$\{S_{q_n}\}_{n=1}^{\infty}$, $S\subset S_{q_n}$, such that each $S_{q_n}$
is a union of $q_n$ intervals, 
$q_n\to\infty$ as $n\to\infty$, and for each $n$:}
\begin{equation}
\left| S_{q_n} \right| < \frac{c}{q_n^{\beta}} ,
\label{lemm1}
\end{equation}
where $\beta$ and $c$ are positive constants; then:
\begin{equation}
D_H(S) \le \frac{1}{1+\beta} .
\label{lemm12}
\end{equation}
{\it Proof}: See Ref.~\cite{la}.\newline

{\bf Theorem~2 :} {\it Assuming Conjecture 1, for any irrational 
$\sigma\in [0,1]$ we have}
\begin{equation}
D_H\big(S(\sigma)\big) \le \frac{1}{2} .
\end{equation}
{\it Proof}: 
Choosing for $S_{q_n}$ in Lemma~1 a cover satisfying Eq.(\ref{corr})
of Corollary~1
and then comparing Eqs.~(\ref{corr}) and (\ref{lemm1})
yields $\beta=1$. Inserting this into Eq.~(\ref{lemm12}) concludes the 
proof of Theorem 2.$\qquad\Box$

\section{The Cover}

In this section we present a specific cover which is numerically shown to 
satisfy Conjecture 1. A good starting point for covering the spectrum of 
Harper's operator [Eq.~(\ref{haop})] for some rational $\sigma'$ 
are the bands of the spectrum for a rational $\sigma$ obtained by 
truncating the continued fraction expansion of $\sigma'$. In fact, there 
is a natural assignment of any of these bands to a cluster of bands of the 
spectrum for $\sigma'$, which is determined by Hofstadter's Rules 
(see Appendix). Very often a band is too narrow to cover its related 
cluster, e.g., the lowest band of $\sigma=1/3$ does not cover the related 
cluster consisting of the lowest two bands for $\sigma'=2/5$. Therefore, 
we introduce the factor 
$R_i(\frac{p}{q},\frac{p'}{q'})$ by which the $i$-th band, 
$i=1,2,\ldots,q$, of the spectrum for $\sigma=p/q$ has to be stretched in 
order to cover its related cluster of the spectrum for $\sigma'=p'/q'$.

We numerically analyzed the stretching factors $R_i$ for all allowed pairs
$p/q$ and $p'/q'$ with $q'\le300$ and found that they are bounded from 
above by
\begin{equation}
\label{maxrf}
\max_{\overset{p,q,p',q',i}{\scriptstyle q'\le 300}}  \, 
R_{i}\left(\frac{p}{q} ,\frac{p'}{q'} \right) \le 2.1 \quad .
\end{equation}
If a bound would exist {\em without} the restriction $q'\le 300$ for any 
two rational approximants $p/q$ and 
$p'/q'$ with $q<q'$, that are obtained by truncating the continued 
fraction expansion of some irrational $\sigma\in[0,1]$, then the cover 
described above would fulfill Conjecture 1. In the remaining part of this 
section we will argue with the help of further numerical analysis that 
this is very likely to be the case.

We first investigate for which bands and field values the maximum 
stretching factor for a fixed $q$ occurs. We observe that the maximum 
stretching factor always occurs when covering the spectrum for $1/(q+1)$ 
with the bands of the spectrum for $1/q$. At these magnetic fields the 
maximum stretching factor stems from the central bands, namely 
\begin{equation}
\label{maxrf1}
\max_{\overset{p,p',q',i}{\scriptstyle q'\le 300}}  \, 
R_{i}\left(\frac{p}{q} ,\frac{p'}{q'} \right) = 
R_{\left\lfloor\frac{q+1}{2}\right\rfloor}\left(\frac{1}{q} ,\frac{1}{q+1}
\right)  ,
\end{equation} 
where $\lfloor x\rfloor$ denotes the integer part of $x$. For example, in 
the case $q=3$ ($q=4$) the maximum stretching factor occurs when using the
second band of $p/q=1/3$ ($p/q=1/4$) to cover the cluster containing the 
second and third band of $p'/q'=1/4$ ($p'/q'=1/5$) (Fig.~1). Figure~2 
shows $R_{\left\lfloor\frac{q+1}{2}\right\rfloor}\left(\frac{1}{q} ,
\frac{1}{q+1} \right)$ as a function of $q$. It has a maximum value of 
about 2.015 at $q\simeq 300$ and is asymptotically decreasing towards 2, 
so that it is bounded from above. This behaviour is substantiated (Fig.~2)
by using results from Thouless and Tan\cite{thtan} giving an approximation
of the band edges for magnetic fields of the form $1/q$. 

Since Hofstadter's butterfly is self-similar, one expects that the 
stretching factors coming from smaller 
distorted copies of the butterfly play an important role.
In fact, one may wonder if they require larger stretching factors, because
of their distortion. For example, such a copy lives between the
lowest two bands of 1/3 and 1/2 (Fig.~1) with an effective magnetic field 
$\sigma_{\rm eff}$ given by $\sigma=1/(2+\sigma_{\rm eff})$.
Assuming that in the smaller copies the same bands as in the original
butterfly need the largest stretching factors, we investigated
the stretching factors of the center bands of this
distorted copy for $\sigma_{\rm eff}=1/q_{\rm eff}$ when covering the 
spectrum for $\sigma'_{\rm eff}=1/(q_{\rm eff}+1)$. As can be seen in 
Fig.~2 (crosses) they converge towards the corresponding stretching 
factors of the original butterfly, however, with deviations for small 
$q_{\rm eff}$. A systematic analysis of these deviations also for 
other small copies of the butterfly shows that these deviations are 
bounded (Fig.~3). This suggests that Eq.~(\ref{maxrf}) remains 
valid even without the restriction $q'\le300$ and thereby gives strong 
numerical evidence for the validity of Conjecture 1.

\section{conclusion}

We have conjectured that the spectrum of the Harper operator
for a rational value of the incommensurability parameter $\sigma$ can be 
covered by the spectrum for a rational approximant of $\sigma$ after 
stretching the bands of the latter by a factor $R$ independent of $\sigma$.
This factor is numerically found to be $R\approx 2.1$. We showed that
from this Conjecture follows that the spectrum is
a zero measure Cantor set for {\it all} irrational $\sigma$. Furthermore, 
it implies that the Hausdorff dimension of these spectra is bounded from 
above by 1/2. While these results are for irrational values of $\sigma$, 
the underlying conjecture is stated for pairs of rational values of 
$\sigma$. This allowed a detailed numerical study giving strong 
numerical evidence for our Conjecture.
As we give an explicit construction of the cover used for the conjecture, 
we hope that our numerical approach is helpful in finding a rigorous proof
of these results.

\acknowledgments

This work was supported in part by the Deutsche Forschungsgemeinschaft. We
thank David Damanik for useful discussions.
R.K. thanks W.~Kohn and D.~Hone for their hospitality while visiting UCSB.

\section*{Appendix}

Here we describe in detail how the cover used in Section IV is constructed
using Hofstadter's rules\cite{hofst,wi1}. 
This set of rules describes qualitatively the self-similar structure of the
graph in Fig.~1. For a given magnetic field $\sigma$ the energy spectrum 
of the Harper operator consists of three non-overlapping parts, one 
central cluster $C$ and two side clusters $L$ and $R$. These three parts 
are distorted Harper spectra for effective field values 
$\sigma_L$, $\sigma_C$ and $\sigma_R$ which are given by
\begin{align}\label{hofreg} \sigma_{L} =\sigma_{R}  &= 
	\begin{cases} 
		\left\{\frac{\displaystyle 1}{\displaystyle \sigma} 
\right\} &\text{for $0<\sigma \le \frac{1}{2}$} \\[2mm]
		\left\{\frac{\displaystyle 1}{\displaystyle 1-\sigma} 
\right\} & \text{for $\frac{1}{2}<\sigma < 1$}  
	\end{cases} \\[2mm]
\sigma_{C} &=  \begin{cases} \left\{\frac{\displaystyle \sigma}
{\displaystyle 1-2\sigma}\right\}& \text{for $0\le\sigma < \frac{1}{2}$} 
\\[2mm]
			  \left\{\frac{\displaystyle 1-\sigma}
{\displaystyle 2\sigma-1}\right\}& \text{for $\frac{1}{2}<\sigma < 1$}
	    \end{cases} \quad ,
\end{align}
where $\{x\}$ denotes the fractional part of $x$. These rules also apply 
to the distorted spectra such that the spectrum splits into three 
clusters, each of which splits into three clusters. For irrational 
$\sigma$ this splitting is continued ad infinitum. In contrast, for 
rational $\sigma$ this process eventually stops for each cluster when it 
consists of either a single band ($\sigma_{\rm eff}=0$) or of two bands 
($\sigma_{\rm eff}=1/2$); see
Fig.~4 for $\sigma=5/8$. For $\sigma'$, where $\sigma$ was obtained by 
truncating the continued fraction 
expansion of $\sigma'$, we simultaneously apply Hofstadter's rules up to 
the same level as for $\sigma$; see
Fig.~4 for $\sigma' = 34/55$. This defines a straightforward assignment of 
a band with $\sigma_{\rm eff}=0$ in the spectrum for $\sigma$ to a cluster
of bands in the spectrum for $\sigma'$. In the case $\sigma_{\rm eff}=1/2$
one has two bands that have to cover the corresponding cluster, which 
consists of three subclusters. We assign one band to two subclusters and 
the other band to the third subcluster (Fig.~4) such that the possibly 
necessary stretching factors for the bands in order to cover their 
subcluster are minimized. By this procedure we assign every band of 
$\sigma$ to a cluster of bands of $\sigma'$, where the union of these 
clusters is the complete spectrum for $\sigma'$.


\begin{figure}
\caption{ The energy spectrum of Harper's operator 
[Eq.(\protect\ref{haop})] for $\lambda=2$ as a function
of the incommensurability parameter $\sigma$. The resulting graph is known
as Hofstadter's butterfly. }
\end{figure}

\begin{figure}
\caption{ Stretching factor 
$R_{\left\lfloor\frac{q+1}{2}\right\rfloor}\left(\frac{1}{q}
,\frac{1}{q+1} \right)$ vs. $q$ (diamonds) compared to its analytic value
(solid line) obtained from an
approximation of the band edges for magnetic fields of the form $1/q$ by 
Thouless and Tan \protect\cite{thtan}. The crosses show the corresponding 
stretching factors of the distorted copy of the butterfly between the 
lowest two bands of 1/3 and 1/2 (Fig.~1) vs. $q_{\rm eff}$ (see text).}
\end{figure}

\begin{figure}
\caption{Maximum deviation $\Delta(\tilde{q})$ of stretching factors in 
the copies of the butterfly around $\tilde{p}/\tilde{q}$ from the 
corresponding stretching factors in the original butterfly (diamonds in 
Fig.~2). Only stretching factors for $1/q_{\rm eff}$, analogous to 
$R_{\left\lfloor\frac{q+1}{2}\right\rfloor}\left(\frac{1}{q},\frac{1}{q+1}
 \right)$ in the original butterfly, are taken into account. For different
$q_{\rm eff}$ the maximum is taken over all copies of the butterfly and 
all $\tilde{p}$. For each $q_{\rm eff}$ the deviations remain bounded and 
the upper bounds decrease with increasing $q_{\rm eff}$.}
\end{figure}

\begin{figure}
\caption{Assignment of bands in the spectrum for $\sigma=5/8$ to clusters 
of bands in the spectrum for $\sigma'=34/55$ by simultaneously applying 
Hofstadter's rules. The assignment is straightforward, except for the case
$\sigma_{\rm eff}=1/2$ where two bands have to be assigned to three 
subclusters (see text). The arrows mark the cases where bands have to be 
stretched in order to cover their assigned cluster.}
\end{figure}

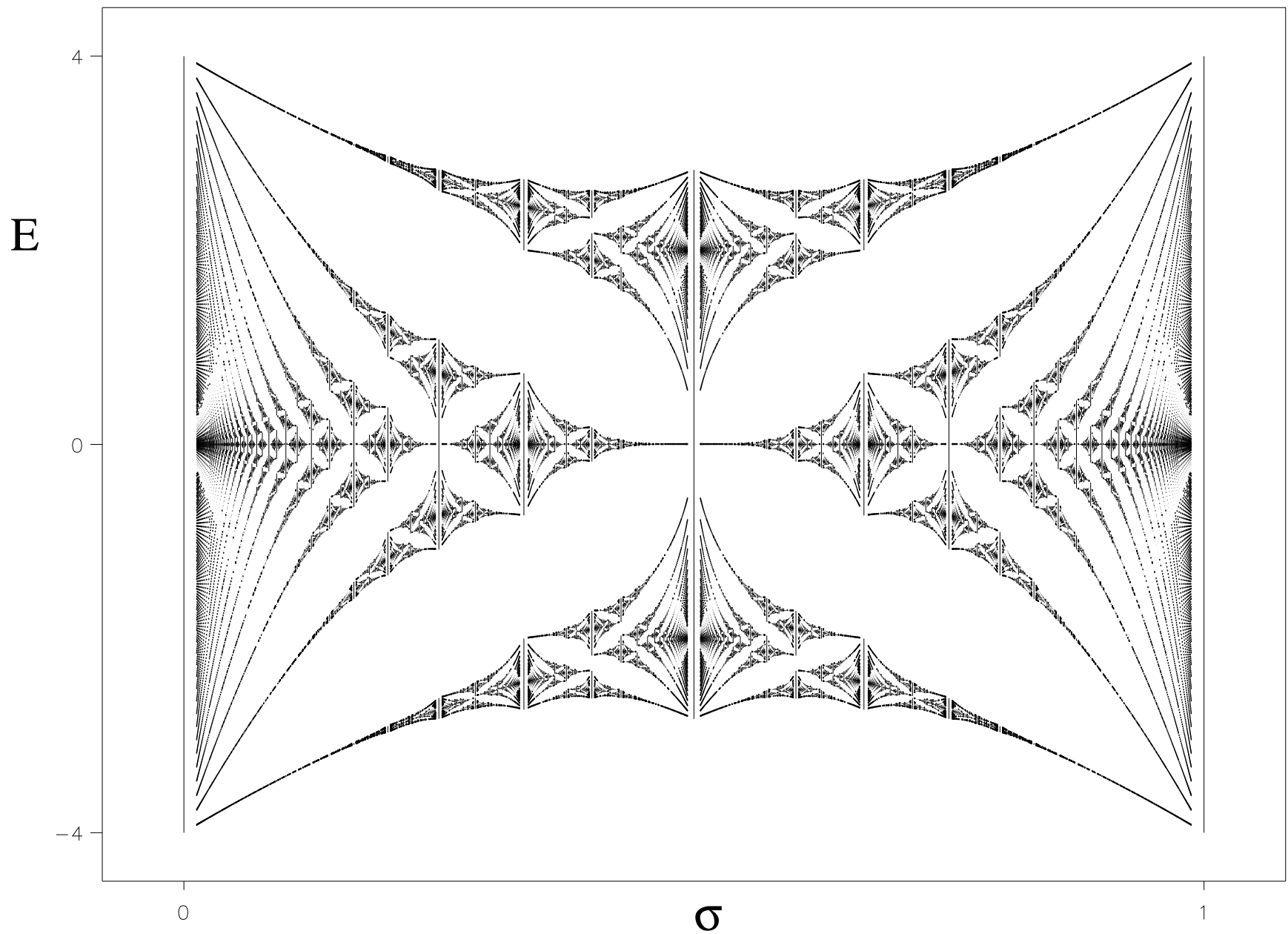
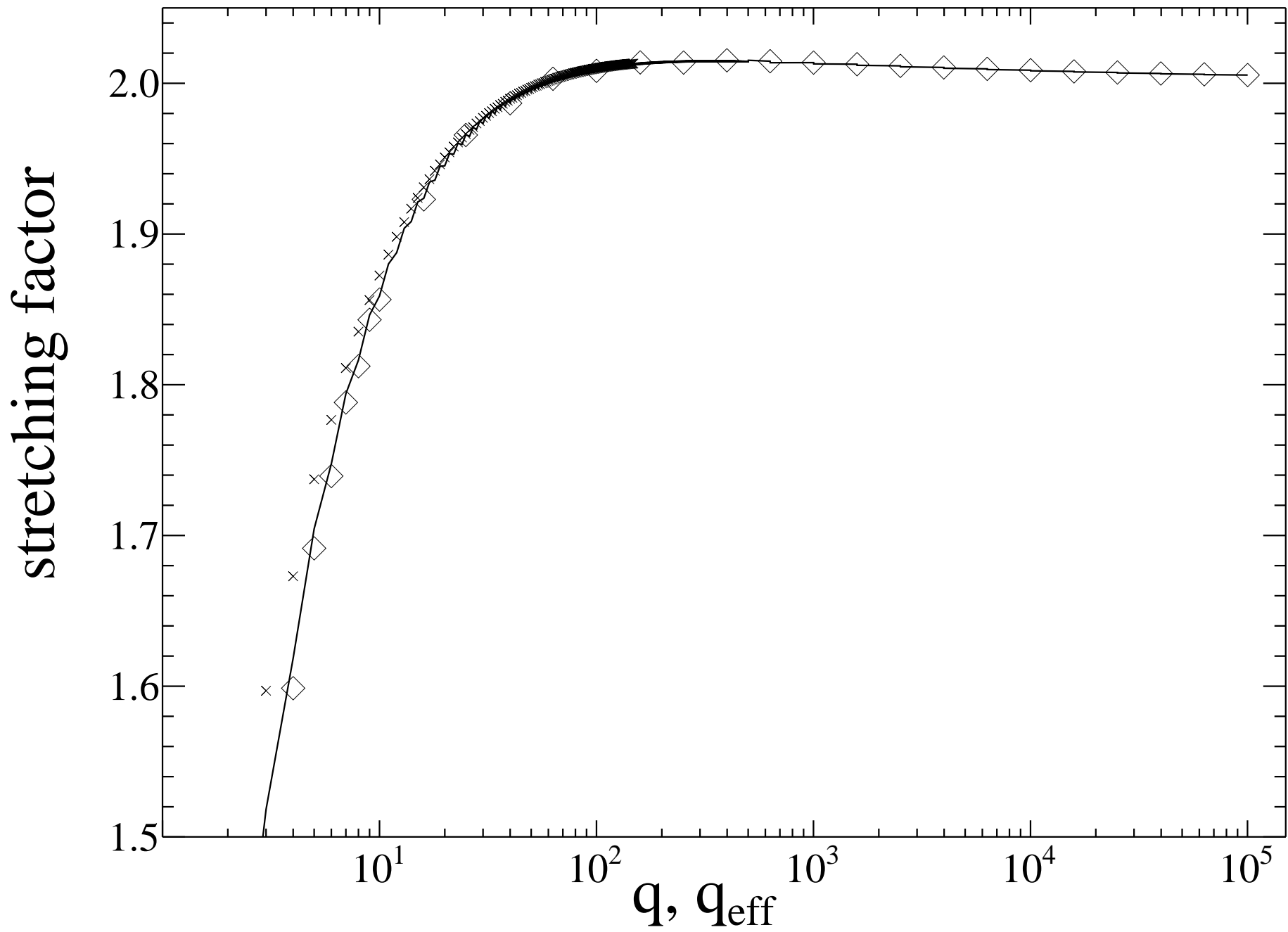
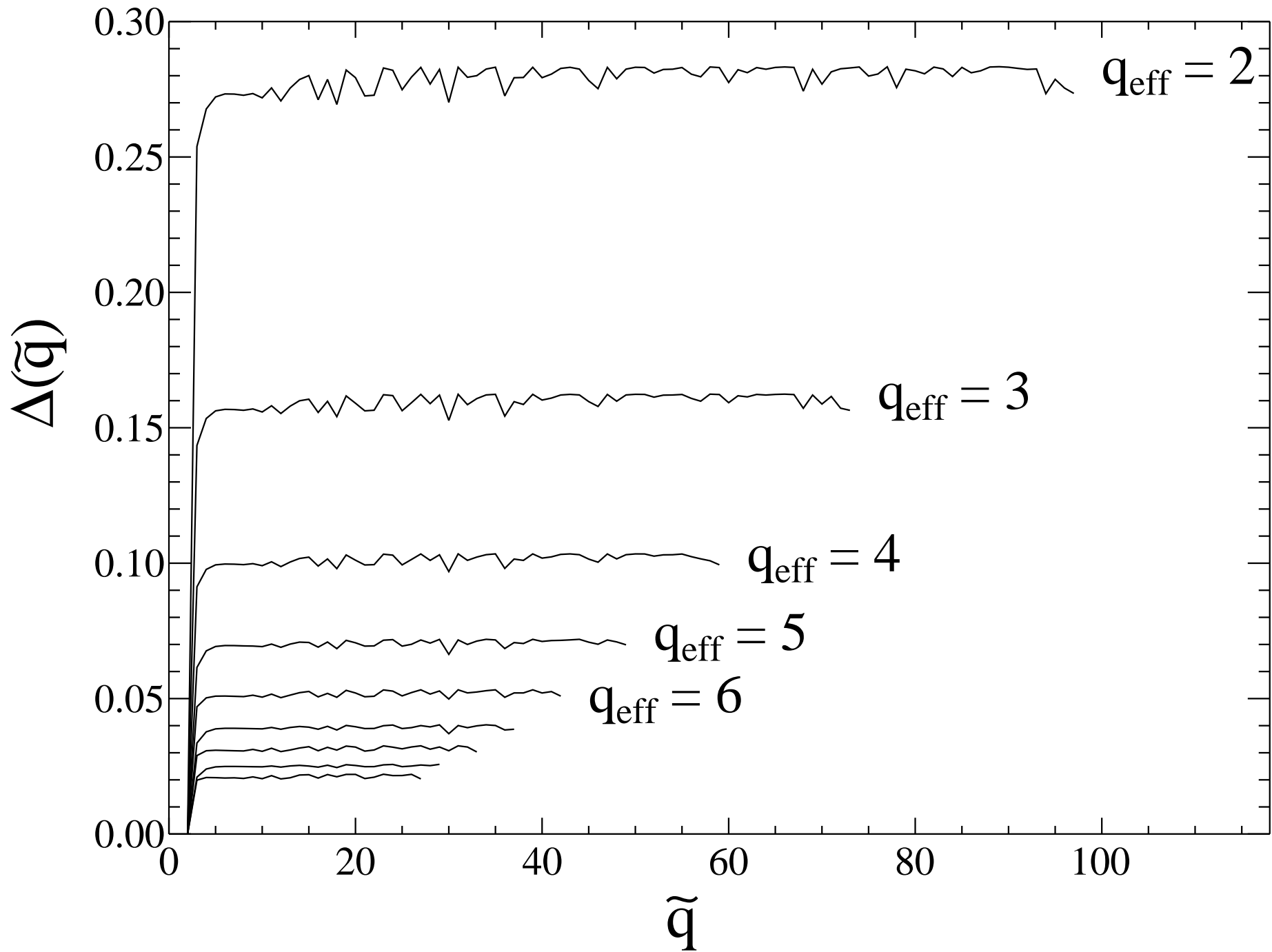
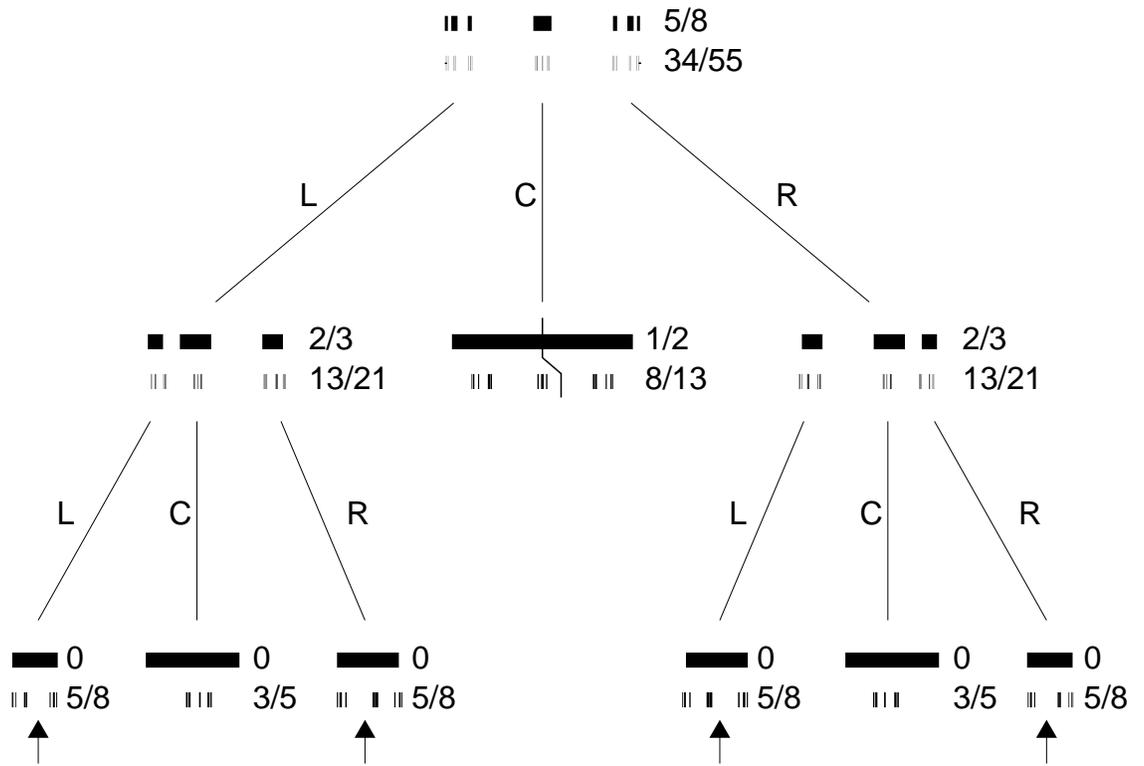
\end{document}